\documentclass[letterpaper, 10 pt, conference]{ieeeconf}  

\IEEEoverridecommandlockouts                              
\usepackage{breakcites}
\usepackage{gensymb}
\usepackage{amsmath} 
\usepackage{amssymb}  
\usepackage[dvipsnames]{xcolor}
\usepackage{makecell}
\usepackage{booktabs}
\usepackage{graphicx}
\usepackage{comment}
\usepackage{subfigure}
\usepackage{cite}
\usepackage{url}

\makeatletter
\let\NAT@parse\undefined
\makeatother
\usepackage{hyperref}
\usepackage{cleveref}
\usepackage{fancyhdr}

\fancypagestyle{firstpage}{\lhead{\vspace{-0.75 cm} \small \textbf{{\fontfamily{cmss}\selectfont
2022 American Control Conference (ACC)\\June 8-10, 2022, Atlanta, GA, USA}}}}

\title{\Large \bf
Development of a Model Predictive Airpath Controller for a Diesel Engine\\ on a High-Fidelity Engine Model with Transient Thermal Dynamics
}

\author{Jiadi Zhang$^{1}$, Mohammad Reza Amini$^{2}$, Ilya Kolmanovsky$^{1}$, Munechika Tsutsumi$^{3}$, and Hayato Nakada$^{3}$ 
\thanks{$^{1}$J. Zhang and I. Kolmanovsky are with the Department of Aerospace Engineering, University of Michigan, Ann Arbor, MI 48109, USA
        {\tt\small \{jiadi,ilya\}@umich.edu}}%
\thanks{$^{2}$M.R. Amini is with the Department of Naval Architecture and Marine Engineering, University of Michigan, Ann Arbor, MI 48109, USA
        {\tt\small mamini@umich.edu}}%
\thanks{$^{3}$H. Nakada and M. Tsutsumi are with Hino Motors, Ltd., Tokyo 191-8660, Japan
       {\tt\small \{mu.tsutsumi,hayato.nakada\}@hino.co.jp}}%
}
\renewcommand{\baselinestretch}{0.85}
\begin{document}

\maketitle
\thispagestyle{firstpage}


\begin{abstract}
\textcolor{black}{This paper presents the results of a model predictive controller (MPC) development for diesel engine air-path regulation. The control objective is to track the intake manifold pressure and exhaust gas recirculation (EGR) rate targets by manipulating the EGR valve and variable geometry turbine (VGT) while satisfying state and control constraints. The MPC controller is designed and verified using a high-fidelity engine model in GT-Power. The controller exploits a low-order rate-based linear parameter-varying (LPV) model for prediction which is identified from transient response data generated by the GT-Power model. It is shown that transient engine thermal dynamics influence the airpath dynamics, specifically the intake manifold pressure response, however, MPC demonstrates robustness against inaccuracies in modeling these thermal dynamics. In particular, we show that MPC can be successfully implemented using a rate-based prediction model with two inputs (EGR and VGT positions) identified from data with steady-state wall temperature dynamics, however, closed-loop performance can be improved if a prediction model (i) is identified from data with transient thermal dynamics, and (ii) has the fuel injection rate as extra model input. Further, the MPC calibration process across the engine operating range to achieve improved performance is addressed. As the MPC calibration is shown to be sensitive to the operating conditions, a fast calibration process is proposed.}  
\end{abstract}
\vspace{-4pt}
\section{Introduction} \label{sec:intro}\vspace{-2pt}
The paper addresses the development of diesel engine airpath control system based on Model Predictive Control (MPC). The control problem is to coordinate Exhaust Gas Recirculation (EGR) valve and Variable Geometry Turbocharger (VGT) actuators to control intake manifold pressure and EGR rate to the specified target values subject to constraints on both actuators and intake manifold pressure. Modern diesel engines rely on EGR and VGT to reduce oxides of nitrogen (NOx) emissions and turbo-lag, and they are nonlinear multivariable systems that are operated in rapid transients. Control challenges for such diesel engines have been examined for over two decades.

MPC has become of increasing interest for engine and powertrain control due to its ability to synergistically coordinate multiple actuators, satisfy constraints, and streamline the control design and calibration process, see e.g.,~\cite{del2010automotive,ortner2007predictive,stewart2008model}. For diesel engine airpath control, specifically, MPC solutions have been developed in \cite{huang2013towards,moriyasu2019diesel,huang2015nonlinear}.%

With the automotive industry seeking to lessen reliance on physical prototypes, more control development is to be done based on simulation models. In this paper, a high fidelity GT-Power engine simulation model is used as an engine surrogate for controller development and verification. For this setting, we confirm the efficacy of the framework described in \cite{liao2020model}, that relies on the identified low-order Linear Parameter Varying (LPV) prediction model in the input-output form, and rate-based MPC formulation. We successfully implement and demonstrate this framework on the GT-Power engine model for a different, larger engine than in \cite{liao2020model}. In the development of our MPC controller, we examine several choices for the prediction model and data used for identification. Specifically, we consider prediction models with two inputs (EGR valve position and VGT position) and three inputs (EGR valve position, VGT position, and fuel injection rate).

Furthermore, we examine the impact of the cylinder and intake manifold wall temperature transient dynamics on the airpath prediction model and closed-loop performance. For this, we consider identifying the LPV model from data obtained from the GT-Power~\cite{GTmanual} engine model with temperatures assumed to be in steady-state (as functions of other engine variables) and with the actual transient temperature dynamics. The former option~
allows to emulate several practical development scenarios, e.g., when data are collected from an engine dynamometer where thermal dynamics are dissimilar from the ones in the vehicle or when a dynamic mean-value engine model calibrated from steady-state engine data is used as an engine surrogate.~Our results indicate that the controller can be successfully designed based on either choice of data, however, improved performance is obtained when the LPV model is identified from data with the transient temperature dynamics.

Finally, we find that the controller can be successfully developed without introducing nonlinear terms in the LPV prediction model; the latter approach was used in \cite{huang2018toward} to improve prediction model accuracy but at the cost of the problem becoming a nonlinear MPC problem.
As a result, our framework allows MPC controller implementation based on quadratic programming and does not require an observer as model states are measured/estimated outputs (intake manifold pressure is measured and EGR rate is already estimated by the nominal strategy).

\vspace{-2pt}
\section{Diesel Engine Airpath Modeling}\label{sec:modeling}
\subsection{High-Fidelity Diesel Engine Model}
\begin{figure}[h!]
\centering
\includegraphics[width=0.4\textwidth]{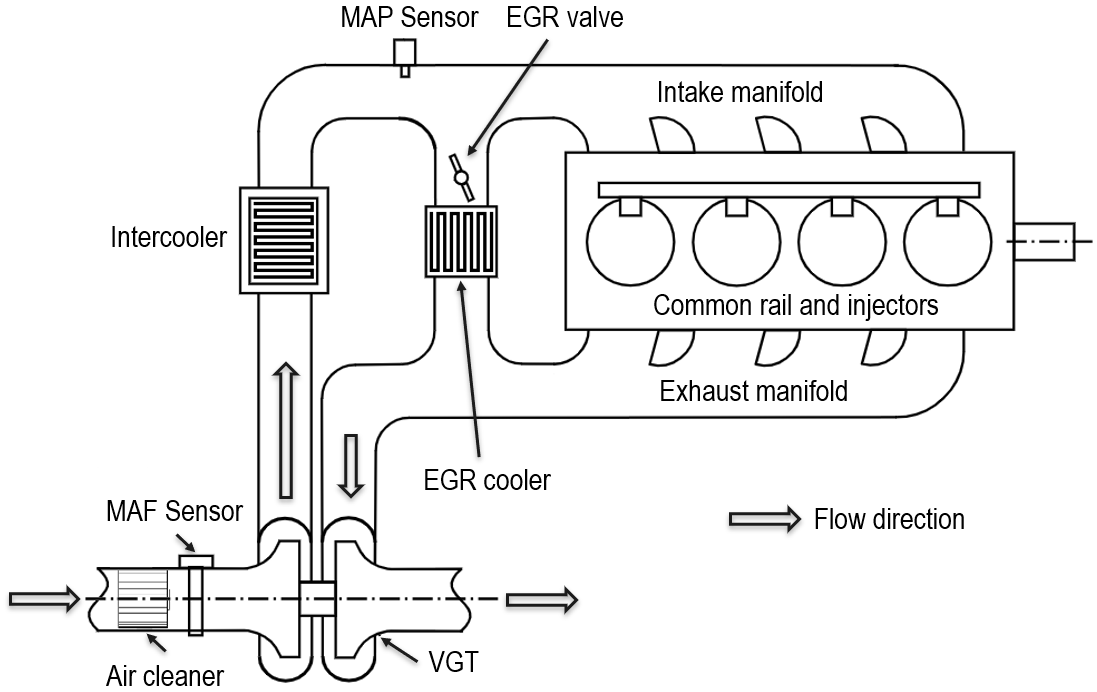}\vspace{-8pt}
\caption{Schematic of a diesel engine and its airpath system.}\vspace{-8pt}
\label{fig:engine}
\end{figure}

\textcolor{black}{
In this paper, a high-fidelity GT-Power engine model is used for the development and verification of MPC controller in simulations.  The real-time capable engine model adopted in this paper represents a diesel engine with details shown in Fig.~\ref{fig:engine}. The engine has four cylinders, intake and exhaust manifolds, an external EGR system and a turbocharger. There are manifold absolute pressure (MAP) and mass air flow (MAF) sensors installed to measure the intake manifold pressure and compressor flow, respectively. 
}

\textcolor{black}{The fuel injection rate is defined as the sum of the pre- and main fuel injection rates. The EGR valve controls the fluid flow rate to the intake manifold and variable geometry turbocharger (VGT) controls the intake manifold pressure by varying the amount of energy extracted from the exhaust gas, which will influence the engine output power. We assume that a set-point strategy or a supervisory controller as in \cite{liao2020model} calculates the set-points for intake manifold pressure and EGR rate at each operating point to balance the output torque and emissions, where the EGR rate is defined as}\vspace{-1pt}
\begin{equation} \label{eq:EGR_rate_def}
\chi_{egr} = {w_{egr}}/({w_{egr}+w_{thr}}),
\end{equation}
where $w_{egr}$ is the mass flow through the EGR valve into the intake manifold and $w_{thr}$ is the throttle mass flow. The airpath controller is designed to track the target intake manifold pressure and EGR rate. Here, we assume target intake manifold pressure and EGR rate are known a priori and can be interpolated using look-up tables provided by OEM. The EGR flow and EGR rate in \eqref{eq:EGR_rate_def} are estimated by the nominal strategy.
\vspace{-5pt}
 \vspace{-5pt}

\subsection{Control-Oriented Diesel Airpath Model} \label{subsec:control-oriented model}
\textcolor{black}{This section describes the development of the control-oriented LPV model to be used as a prediction model in MPC of the diesel engine airpath. To keep the model as simple as possible, intake manifold pressure ($p_{im}$) and EGR rate ($\chi_{egr}$) are selected as the only states of the system ($x$) representing the airpath dynamics~\cite{huang2018toward}.}~\textcolor{black}{The model inputs $u$ are EGR valve position (percent open) and VGT position (percent close). As the intake manifold pressure is measured and EGR rate is estimated, this model structure eliminates the need for an observer.~Given that the dynamics of the engine airpath are nonlinear, we develop linear models for different operating points of the engine, forming an LPV model. The engine operating points $\rho$ are defined by the combination of the engine speed $N_e$ and total fuel injection rate $w_{inj}$.}

Linear models at different operating points are identified from data obtained by perturbing GT-Power model inputs ($u$) by random steps of about 10\% of steady-state magnitudes. \textcolor{black}{Thus, the model identified for the operating point $\rho_i, i=1,\cdots,99$ has the following form,}
\vspace{-4pt}
\begin{equation}  \label{eq:LPV model}
x_{k+1} - x_{k+1}^{ss} = A_i (x_k-x_k^{ss}) + B_i (u_k-u_k^{ss}),
\end{equation}
\textcolor{black}{where $A_i$ and $B_i$ are the matrices identified at the operating point $\rho_i$ with corresponding state ($x^{ss}$) and input ($u^{ss}$) at steady-state.} To identify $A_i$ and $B_i$ in \eqref{eq:LPV model}, {\tt Parameter Estimation} toolbox in MATLAB/Simulink is used. {The training data are generated from the open-loop GT-Power simulations at each $\rho_i$ while} {the validation data are generated from a different set of input perturbations.}

\textcolor{black}{To construct the LPV model, the elements of $A_i$ and $B_i$ are linearly interpolated between the 99 operating points as a function of engine speed and fuel injection rate. Since the look-up table is relatively large ($11\times9$) and the operating points are close enough, we use linear interpolation to estimate the airpath dynamics between two adjacent operating points. The final LPV model has the following form, }
\begin{equation} \label{eq:model}
\begin{split}
x_{k+1} - x_{k+1}^{ss}(\rho_k) = & A(\rho_k) [x_k-x_k^{ss}(\rho_k)] \\
& + B(\rho_k) [u_k-u_k^{ss}(\rho_k)],
\end{split}
\end{equation}
\textcolor{black}{where $A,B: \mathbb{R}^2 \to \mathbb{R}^{2\times2}$ and $x_k^{ss}, u_k^{ss}: \mathbb{R}^2 \to \mathbb{R}^2$. }

\begin{figure}[h!]
\centering
\includegraphics[width=0.8\columnwidth]{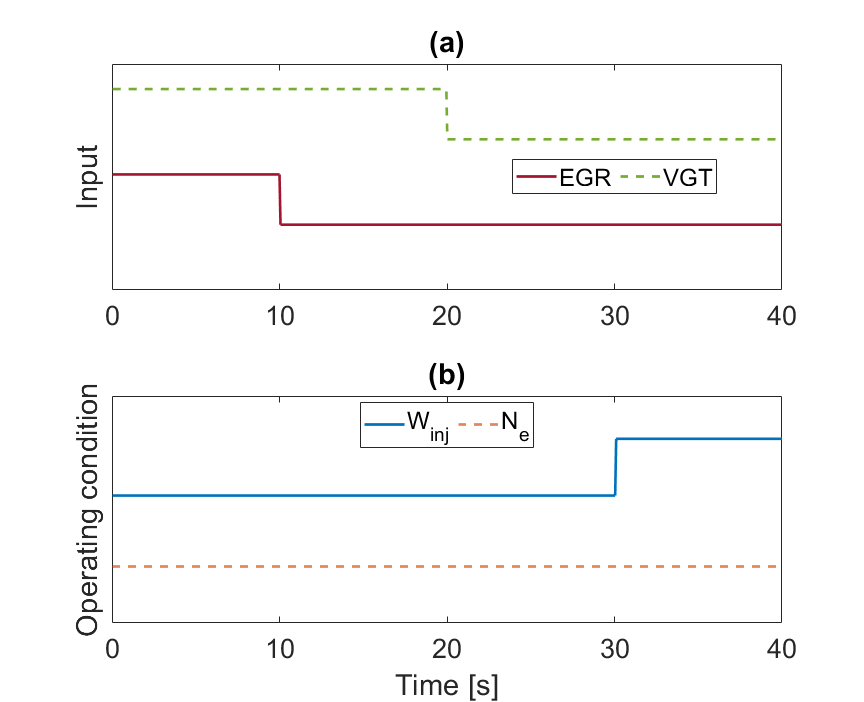}\vspace{-8pt}
\caption{Step inputs and operating conditions applied to the GT-Power model to generate the airpath response of the engine as shown in Fig.~\ref{fig:step comparison}: (a) EGR and VGT positions, and (b) fuel injection rate and engine speed.}\vspace{-16pt}
\label{fig:step thermal dynamic input}
\end{figure}

\begin{figure}[h!]
\centering
\includegraphics[width=0.8\columnwidth]{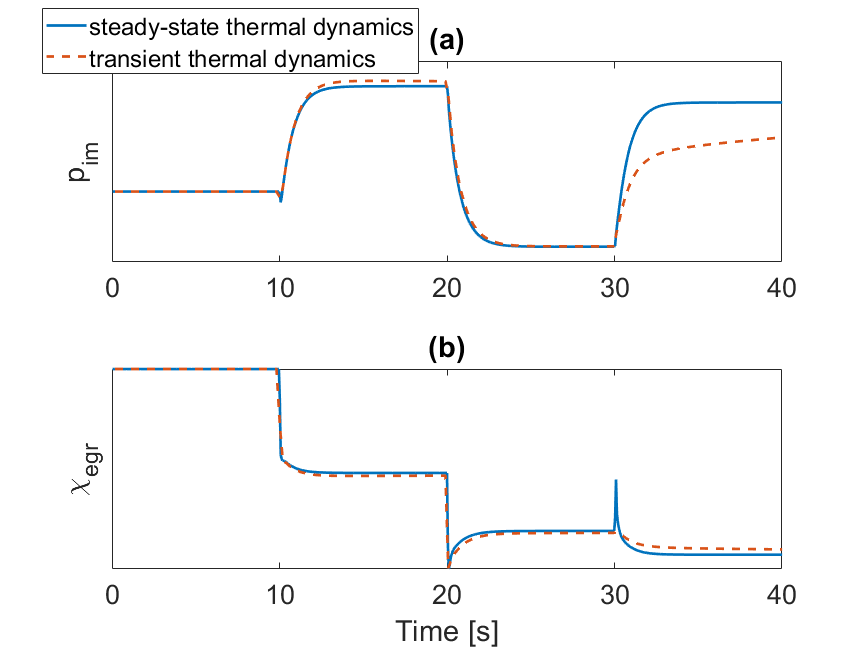}\vspace{-8pt}
\caption{Comparison of the GT-Power model response to step inputs shown in Fig.~\ref{fig:step thermal dynamic input} with assumed steady-state and transient thermal dynamics: (a) intake manifold pressure, and (b) EGR rate.}\vspace{-8pt}
\label{fig:step comparison}
\end{figure}

Two different sets of models \eqref{eq:model} were identified from data generated for two cases. The first case (\textbf{Case I}) is based on the assumption that the thermal dynamics, such as the cylinder wall temperature and intake manifold temperature, have steady-state behavior, i.e., at a given engine speed and fuel injection rate, the temperatures remain at steady-state values. For the second case (\textbf{Case II}), the transient responses of the thermal dynamics are taken into account even if the engine operating condition remains the same. The $p_{im}$ and $\chi_{egr}$ trajectories from Case I and Case II after applying the inputs shown in Fig. \ref{fig:step thermal dynamic input} are recorded and plotted in Fig.~\ref{fig:step comparison}, demonstrating the effect of thermal dynamics on engine air-path responses. Fig.~\ref{fig:step comparison} shows that the airpath responses with transient and steady-state thermal dynamics are similar during a constant operating condition. However, when the operating condition changes, the airpath response with transient thermal dynamics, especially $p_{im}$, will react much slower than that with steady-state thermal dynamics. To investigate how the thermal dynamics influence the LPV model accuracy,~LPV models are developed based on three different strategies:
\begin{itemize}
  \item \textbf{Model A}: develop LPV model based on data from GT-Power model with steady-state thermal dynamics
  \item \textbf{Model B}: develop LPV model based on data from GT-Power model with transient thermal dynamics
  \item \textbf{Model C}: develop LPV model based on data from GT-Power model with transient thermal dynamics and fuel injection rate ($w_{inj}$) as an extra input
\end{itemize}

\vspace{-13pt}

\begin{figure}[t]
\centering
\includegraphics[width=0.8\columnwidth]{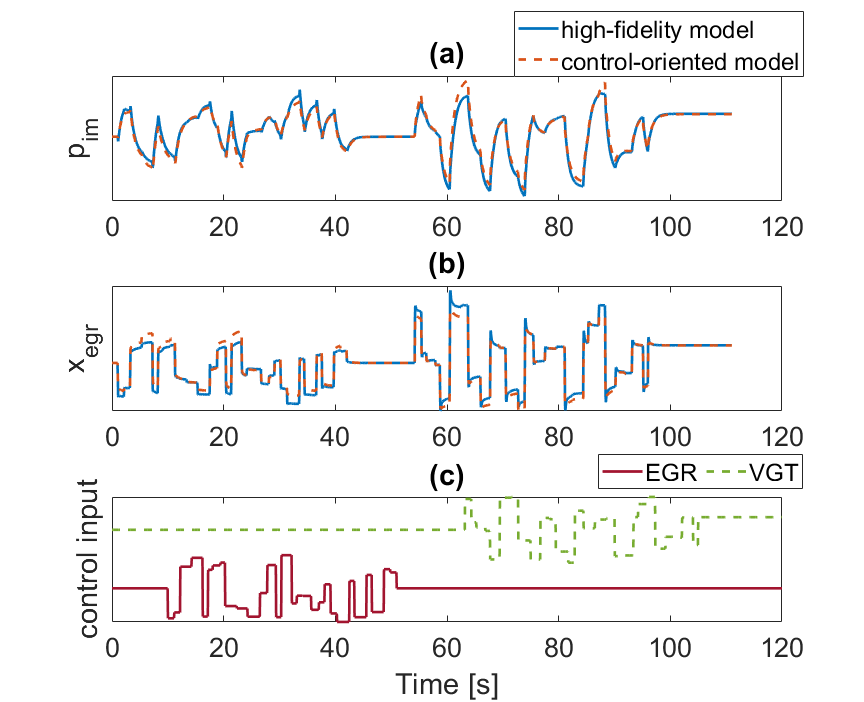}\vspace{-6pt}
\caption{{Validation results of the control-oriented LPV \textit{Model-A} at $2000\ \rm rpm$ and $60\ \rm mm^3/cyl$ based on engine I/O data collected from GT-Power model with steady-state thermal dynamics: (a) intake pressure, (b) EGR rate, and (c) perturbed model inputs for identification.}}\vspace{-13pt}
\label{fig:SID results 1}
\end{figure}

To develop the initial LPV model (\textit{Model-A}),~the thermal wall solver of the GT-Power model for pipe/flowsplit walls and other thermal masses is configured to only calculate the steady-state temperature and the thermal capacitance is not considered \cite{GTmanual}. Fig.~\ref{fig:SID results 1} shows sample validation results of \textit{Model-A} at engine speed of $2000\ \rm rpm$ and fuel injection rate of $60\ \rm mm^3/cyl$. The mean model error for $p_{im}$ and $\chi_{egr}$ are 0.0045 $bar$ and 0.0024, respectively. As can be seen, at this particular operating point, \textit{Model-A} is able to predict the trends in intake manifold pressure and EGR rate.~Note that the numerical values are removed from the axes to preserve data confidentiality.

\begin{figure}[h!]
\centering
\includegraphics[width=0.85\columnwidth]{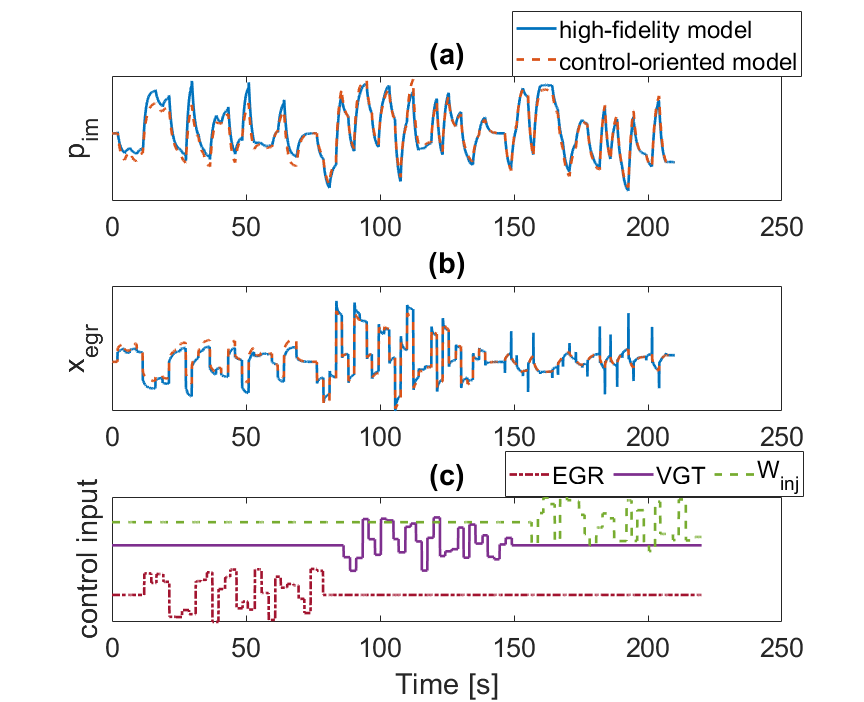}\vspace{-6pt}
\caption{Validation results of \textit{Model-C} at $2000\ \rm rpm$ and $60\ \rm mm^3/cyl$ based on engine I/O data collected from GT-Power model with transient thermal dynamics: (a) intake pressure, (b) EGR rate, (c) perturbed model inputs for identification.}\vspace{-16pt}
\label{fig:SID results 2}
\end{figure}

The second and third LPV models (\textit{Model-B} and \textit{Model-C}) are identified from the GT-Power model with transient thermal dynamics.~Compared with \textit{Model-B}, \textit{Model-C} has fuel injection rate ($w_{inj}$) as an extra input, which is included to better model the slow airpath responses when the fuel injection rate changes as shown in Fig.~\ref{fig:step comparison} from $t = 30~s$ to $40~s$. Inclusion of $w_{inj}$, not only changes $B(\rho_k)$ in \eqref{eq:model} from $\mathbb{R}^{2\times2}$ to $\mathbb{R}^{2\times3}$, it also affects other elements of $A_i$ and $B_i$. Thereby, $A_i$ and $B_i$ need to be re-identified. Fig.~\ref{fig:SID results 2} shows the validation example of \textit{Model-C} with three inputs at $2000\ \rm rpm$ and $60\ \rm mm^3/cyl$.
\vspace{-10pt}

\subsection{Validation of Control-Oriented LPV Model}

To test the performance of LPV \textit{Model-A}, \textit{Model-B}, and \textit{Model-C} across various engine operating regions, they are simulated over the Federal Test Procedure (FTP) driving cycle. 
The GT-Power model with transient thermal wall solver, as the ground truth, is also simulated over the same driving cycle. The inputs to the GT-Power model are first determined according to $p_{im}$ and $\chi_{egr}$ set-point look-up tables provided by the OEM. In the absence of a closed-loop controller, a $\pm5\%$ random perturbation for both EGR and VGT signals is applied at each time step to emulate the closed-loop control signals.~The same set of inputs and operating conditions are used to run all the three models in \cref{subsec:control-oriented model} initialized with the same boundary conditions as the GT-Power model.

\begin{figure}[h!]
\centering
\includegraphics[width=0.85\columnwidth]{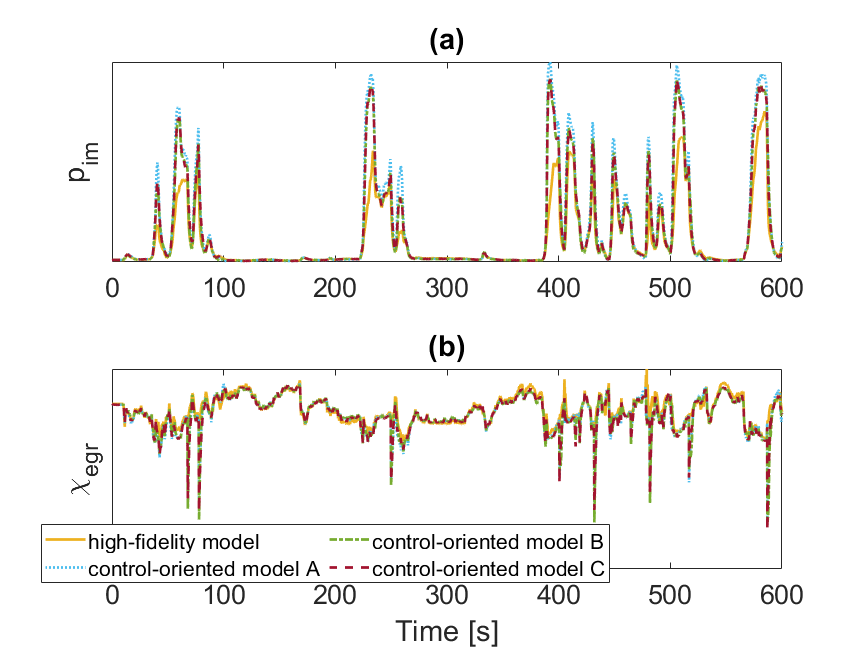}\vspace{-6pt}
\caption{Comparison of LPV \textit{Model-A}, \textit{Model-B} and \textit{Model-C} against the high-fidelity GT-Power model simulated over the FTP driving cycle: (a) intake manifold pressure, and (b) EGR rate.}\vspace{-12pt}
\label{fig:FTP validation}
\end{figure}

The performance of LPV models over the FTP cycle in predicting the engine airpath states are shown in Fig.~\ref{fig:FTP validation}, and compared against the high-fidelity GT-Power model. The results are also summarized in Table \ref{tbl:model}. 
\vspace{-10pt}
\vspace{-5pt}
\begin{table}[h!]
\caption{Mean and standard deviation (STD) of the absolute error for three model validation results over the FTP cycle.}\vspace{-10pt}
\label{tbl:model}
{\scriptsize
\begin{center}
\begin{tabular}{lllll}
\toprule  
\makecell[l]{\textbf{Model}} & \makecell[l]{\textbf{ $\overline{e}_{p_{im}}$}[bar]} & \makecell[l]{\textbf{$\overline{e}_{\chi_{egr}}$}} & \makecell[l]{\textbf{STD: $e_{p_{im}}$}[bar]} & \makecell[l]{\textbf{STD: $e_{\chi_{egr}}$}}\\
\midrule  
\textit{Model-A} & 0.1248 & 0.0256 & 0.2579 & 0.0448\\ 
(reference) &  &  &  & \\ \hline
\textit{Model-B}  & 0.0873 & 0.0250 & 0.2057 & 0.0489\\
         & ($\downarrow$ 30\%) & ($\downarrow$ 2.3\%) & ($\downarrow$ 20.2\%) & ($\uparrow$ 9.2\%)\\ \hline
\textit{Model-C} & 0.0884 & 0.0241 & 0.2061 & 0.0456\\
        & ($\downarrow$ 29.1\%) & ($\downarrow$ 5.8\%) & ($\downarrow$ 20.1\%) & ($\uparrow$ 1.8\%)\\
\bottomrule 
\end{tabular}\vspace{-10pt}
\end{center}}
\end{table}

According to Table \ref{tbl:model}, the average $p_{im}$ and $\chi_{egr}$ prediction errors of \textit{Model-B} are observed to be 30\% and 2.3\% less than \textit{Model-A}, respectively. This shows that the model identified from data with transient thermal behavior could lead to better prediction results. Moreover, compared with \textit{Model-A}, \textit{Model-C} has 29.1\% and 5.8\% less $p_{im}$ and $\chi_{egr}$ prediction errors, respectively. By comparing \textit{Model-B} and \textit{Model-C}, it can be observed that \textit{Model-C} leads to slightly better $\chi_{egr}$ prediction error by 3.5\% on average. As compared to \textit{Model-A}, while \textit{Model-B} and \textit{Model-C} lead to better $\chi_{egr}$ prediction errors, the standard deviation of their prediction errors increase. This observation may suggest that the use of \textit{Model-B} and \textit{Model-C} does not necessarily lead to better $\chi_{egr}$ predictions. For intake manifold pressure, on the other hand, the results in Table \ref{tbl:model} indicate that $p_{im}$ model is sensitive to engine thermal dynamics. Such observation is consistent with the case study reported in Fig.~\ref{fig:step comparison}. By taking into account the engine transient thermal response, either through \textit{Model-B} or \textit{Model-C}, the average prediction errors for $p_{im}$ from \textit{Model-B} and \textit{Model-C} decreases by 30\% and 29.1\%, respectively, as compared to \textit{Model-A}.

It was demonstrated in this section that the diesel engine airpath dynamics, specifically, the intake manifold pressure, are sensitive to transient engine thermal dynamics. It was also shown that to capture such sensitivity by a control-oriented LPV model, it is beneficial to include (i) engine I/O data with transient thermal responses (\textit{Model-B}), and (ii) fuel injection rate (\textit{Model-C}) in the model identification process. Having a more accurate model for predicting the airpath dynamics will facilitate the design of a closed-loop controller to regulate intake manifold pressure and EGR rate. Nevertheless, one key feature of closed-loop control systems is the inherent robustness through the feedback mechanism. Given the slow dynamics of the engine thermal systems, the robustness gained through a closed-loop control policy may compensate for the impact of thermal dynamics on the airpath system, reducing the need for a more accurate LPV model. To investigate the closed-loop control system of the airpath system and its sensitivity to engine transient thermal dynamics, the developed LPV models in this section are next used to develop MPCs in the next section.

\vspace{-3pt}
\section{MPC for Diesel Airpath based on LPV models}\label{sec:control design}
\begin{figure}[h!]
\centering
\includegraphics[width=0.99\columnwidth]{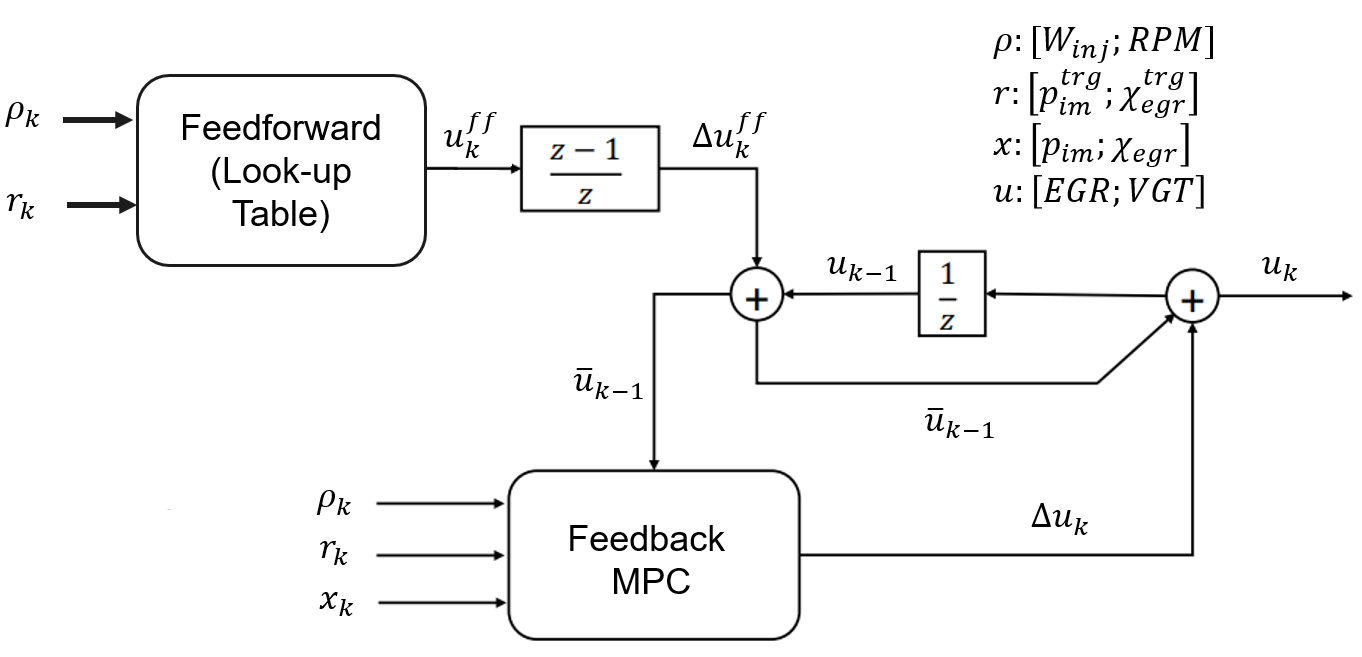}\vspace{-6pt}
\caption{Schematic of airpath MPC architecture. In the square box $z$ stands for the discrete operator, i.e., $u_{k+1} = zu_k$. Note that $\Delta u$ is computed from the feedback MPC, $u^{ff}$ is interpolated from the feedforward look-up table, $u$ is applied to the engine (GT-Power model), and $\bar{u}$ is supplied to the feedback controller through \eqref{subeq:u}.}\vspace{-6pt}
\label{fig:MPC structure}
\end{figure}

The architecture of the employed airpath control system is a combination of feedforward and feedback\cite{liao2020model} controllers shown in Fig.~\ref{fig:MPC structure}. The feedforward controller is incorporated to speed up the airpath dynamic responses. Unlike \cite{liao2020model}, here we use a look-up table rather than an MPC loop for the feedforward ($u^{ff}$), \textcolor{black}{with respect to $p_{\tt im}$ and $\chi_{\tt egr}$ set-points at given engine speed and fuel injection rate values in steady-state.}~The feedback controller, on the other hand, is implemented for enhanced robustness and disturbance rejection.
\vspace{-10pt}

In this paper, a linear MPC is used for the feedback loop. Due to plant-model mismatch, integral action is needed to achieve zero offset steady-state tracking. One way to incorporate integral action is using a rate-based MPC~\cite{wang2004tutorial,pannocchia2015offset}. With such a method, the rate-based model\vspace{-2pt}
\begin{equation} \label{eq:rate base model}
\Delta x_{k+1} = A(\rho_k) \Delta x_k + B(\rho_k) \Delta u_k,
\end{equation}
is used, where $\Delta x_{k} = x_{k} - x_{k-1}$ and $\Delta u_{k} = u_{k} - u_{k-1}$. The rate-based prediction model \eqref{eq:rate base model} is synergistic with the employed LPV framework, as the steady-state values of states and controls in \eqref{eq:model} do not need to be known, assuming they remain constant over the prediction horizon.~The feedforward control signal is interpolated from a look-up table and added to the integral action in the feedback loop, generating the final control signal applied to the system.

\textcolor{black}{Based on the rate-based model \eqref{eq:rate base model}, the MPC is formulated with the following cost function},
\vspace{-6pt}
\begin{subequations}\label{eq:origional MPC}
\begin{equation} 
\min_{\Delta u_{0|k},...,\Delta u_{N-1|k}} \sum_{j=0}^{N-1} e_{j|k}^TQ_{e}e_{j|k} + \Delta u_{j|k}^TR\Delta u_{j|k}
\tag{\ref{eq:origional MPC}}
\end{equation}
subject to 
\vspace{-6pt}
\begin{align} 
&\Delta x_{j+1|k} = A(\rho_k) \Delta x_{j|k} + B(\rho_k) \Delta u_{j|k},\\
&\Delta x_{0|k} = x_{k} - x_{k-1} \\
&e_{0|k} = x_{k} - r_k\\
&e_{j+1|k} = A(\rho_k)\Delta x_{j|k} + B(\rho_k)\Delta u_{j|k} + e_{j|k}\\
&x_{j|k} = x_{j-1|k} + \Delta x_{j|k}, j = 1,.., N,\\
&u_{j|k} = u_{j-1|k} + \Delta u_{j|k}, j = 0,.., N-1, \label{subeq:u}\\
&x_{min} \leq x_{j|k} \leq x_{max}, j = 1,..., N,\\
&u_{min} \leq u_{j|k} \leq u_{max}, j = 0,..., N-1,
\end{align}
\end{subequations}
where $N$ is the prediction horizon and $Q \succeq 0$ and $R \succ 0$ are weighting matrices. The index $j$ runs over the prediction horizon while the index $k$ indicates the sampling instance.~\textcolor{black}{
An augmented model is used in the MPC design} with the state $x_{j|k}^{ext} = [\Delta x_{j|k}^T, e_{j|k}^T, x_{j-1|k}^T, u_{j-1|k}^T]^T$ as extended state vector and considering $\Delta u_{j|k}$ as the control input. To ensure closed-loop stability, a terminal penalty $P_{\infty|k}$ on $\Delta x_{N|k}$ and $e_{N|k}$ is augmented based on the solution to the discrete algebraic Riccati equation (DARE) corresponding to dynamics and input matrices of the form,
\vspace{-5pt}
\begin{equation}
\begin{bmatrix}
A(\rho_k) & 0\\
A(\rho_k) & \mathbb{I}_{n_e\times n_e}
\end{bmatrix}
, 
\begin{bmatrix}
B(\rho_k)\\
B(\rho_k)
\end{bmatrix},
\end{equation}
\textcolor{black}{with state and control weighting matrices selected by the designer,}
\vspace{-5pt}
\begin{equation}
\begin{bmatrix}
0 & 0\\
0 & Q_e
\end{bmatrix}
, 
R.
\end{equation}

\textcolor{black}{A slack variable $\epsilon_k\geq 0$ is also introduced to relax the state constraint as,}
\vspace{-7pt}
\begin{align*}
    &x_{min} - \epsilon_k \leq x_j \leq x_{max} + \epsilon_k,
\end{align*}
\textcolor{black}{which will be weighted in the cost. Here $x_{min}$ and $x_{max}$ are limits on intake manifold pressure and EGR rate. Incorporating all these changes, the final form of the rate-based MPC is,}
%
\vspace{-5pt}
\begin{subequations}\label{eq:modified MPC}
\begin{multline}
    \min_{\Delta u_{0|k},...,\Delta u_{N-1|k},\epsilon_k} (x_{N|k}^{ext})^TP_{N|k}x_{N|k}^{ext} +  \\
    \sum_{j=0}^{N-1} (x_{j|k}^{ext})^TQ^{ext}x_{j|k}^{ext} + \Delta u_{j|k}^TR^{ext}\Delta u_{j|k} + \mu\epsilon_k^T\epsilon_k
\tag{\ref{eq:modified MPC}}
\end{multline}
subject to\vspace{-5pt}
\begin{align} 
    &x_{j+1|k}^{ext} = \begin{bmatrix}
    A(\rho_k) & 0 & 0 & 0\\
    A(\rho_k) & \mathbb{I}_{n_e\times n_e} & 0 & 0\\
    \mathbb{I}_{n_x\times n_x} & 0 & \mathbb{I}_{n_x\times n_x} & 0\\
    0 & 0 & 0 & \mathbb{I}_{n_u\times n_u}
    \end{bmatrix}
    x_{j|k}^{ext} + \nonumber\\
    & \begin{bmatrix}
    B(\rho_k) \\
    B(\rho_k) \\
    0\\
    \mathbb{I}_{n_u\times n_u}
    \end{bmatrix} \Delta u_{j|k},\\
    &\Delta x_{0|k} = x_{k} - x_{k-1} \\
    &e_{0|k} = x_{k} - r_k\\
    &x_{-1|k} = x_{k-1}\\
    &u_{-1|k} = \bar{u}_{k-1}\\
    &e_{j+1|k} = A(\rho_k)\Delta x_{j|k} + B(\rho_k)\Delta u_{j|k} + e_{j|k}\\
    &x_{j|k} = x_{j-1|k} + \Delta x_{j|k}, j = 1,.., N,\\
    &u_{j|k} = u_{j-1|k} + \Delta u_{j|k}, j = 0,.., N-1,\\
    &x_{min} - \epsilon_k\leq x_{j|k} \leq x_{max} + \epsilon_k, j = 1,..., N,\\
    &u_{min} \leq u_{j|k} \leq u_{max}, j = 0,..., N-1,
\end{align}
\end{subequations}
where $\bar{u}_{k-1} = u_{k-1} + u_k^{ff} - u_{k-1}^{ff}$.

\textcolor{black}{Over the prediction horizon, the operating condition is treated as fixed, which means for the model introduced in \cref{subsec:control-oriented model} with 3-D input, $\Delta w_{inj} = 0$ throughout the prediction horizon. Thus, the third input for the MPC implementation can be ignored during optimization and the $B(\rho_k)$ could be reduced from $\mathbb{R}^{2\times3}$ to $\mathbb{R}^{2\times2}$ by eliminating the last column of $B(\rho_k)$.} 

\vspace{-4pt}
\section{Airpath MPC Calibration}\label{sec:control implementation}
\begin{figure}[h!]
\centering
\includegraphics[width=0.8\columnwidth]{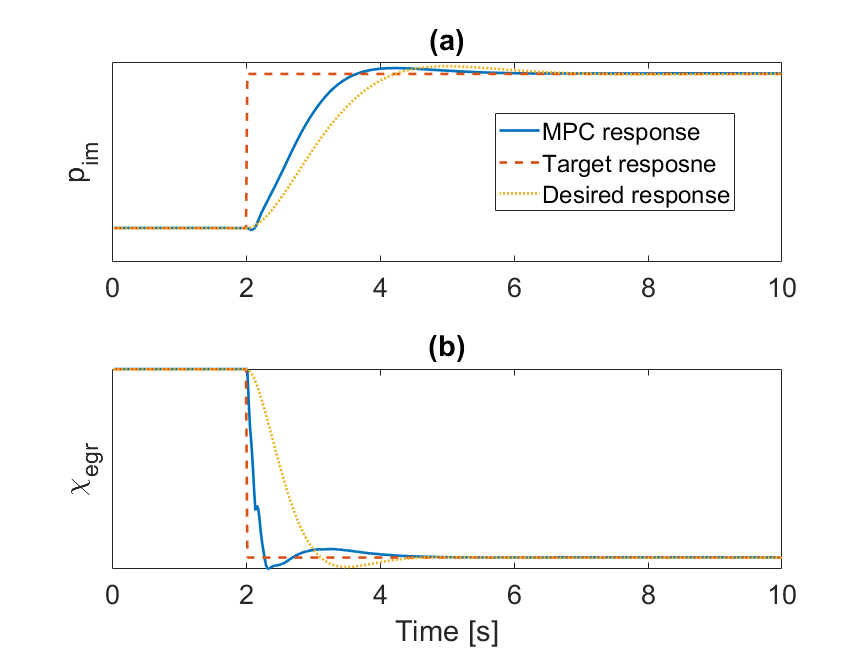}\vspace{-6pt}
\caption{Example of the MPC tuning process with \textit{Model-A} at $1000\ \rm rpm$ and $40\ \rm mm^3/cyl$: (a) intake manifold pressure, and (b) EGR rate.}\vspace{-8pt}
\label{fig:tuning example}
\end{figure}

\vspace{-3pt}
\textcolor{black}{Implementation of diesel airpath MPC requires the selection of weighting matrices $Q^{ext}$ and $R^{ext}$ in \eqref{eq:modified MPC}. Our investigation shows that it is beneficial to select $Q^{ext}$ and $R^{ext}$ differently for different engine operating conditions. This is due to the highly nonlinear dynamics of the engine and the difference between the linear sub-models developed at different operating points.~Here, we develop a framework for tuning of $Q^{ext}$ and $R^{ext}$ pair at each engine operating point.} First, for $p_{im}$, we set the desired maximum 90 percentile response time (referred to as ``response time'' in the rest of the paper) and overshoot $OS_{p_{im}}$, which is modeled using a second-order system as
\vspace{-4pt}
\begin{equation}\label{eq:second order system}
    p_{im}^{desired} = G(z)p_{im}^{trg},\vspace{-4pt}
\end{equation}
where $G(z)$ is a discrete second order transfer function with the time constant $\tau_{p_{im}}$ and the overshoot $OS_{p_{im}}$. Parameters of \eqref{eq:second order system} can be chosen by benchmarking another engine or requirement cascade process. Next, we apply a fuel step from the nominal fuel set-point and check if the response time of $p_{im}$ is faster than the desired value (see the yellow dot line in Fig.~\ref{fig:tuning example}-(a)). After reaching the desired response or better, the process is repeated for $\chi_{egr}$ tuning (see Fig.\ref{fig:tuning example}-(b)). Finally, the results are checked to ensure both $p_{im}$ and $\chi_{egr}$ demonstrate desirable performance. Otherwise, the process is repeated from the beginning until the desired response is reached by making further adjustments.
\vspace{-11pt}

\begin{figure}[t]
\centering
\includegraphics[width=0.9\columnwidth]{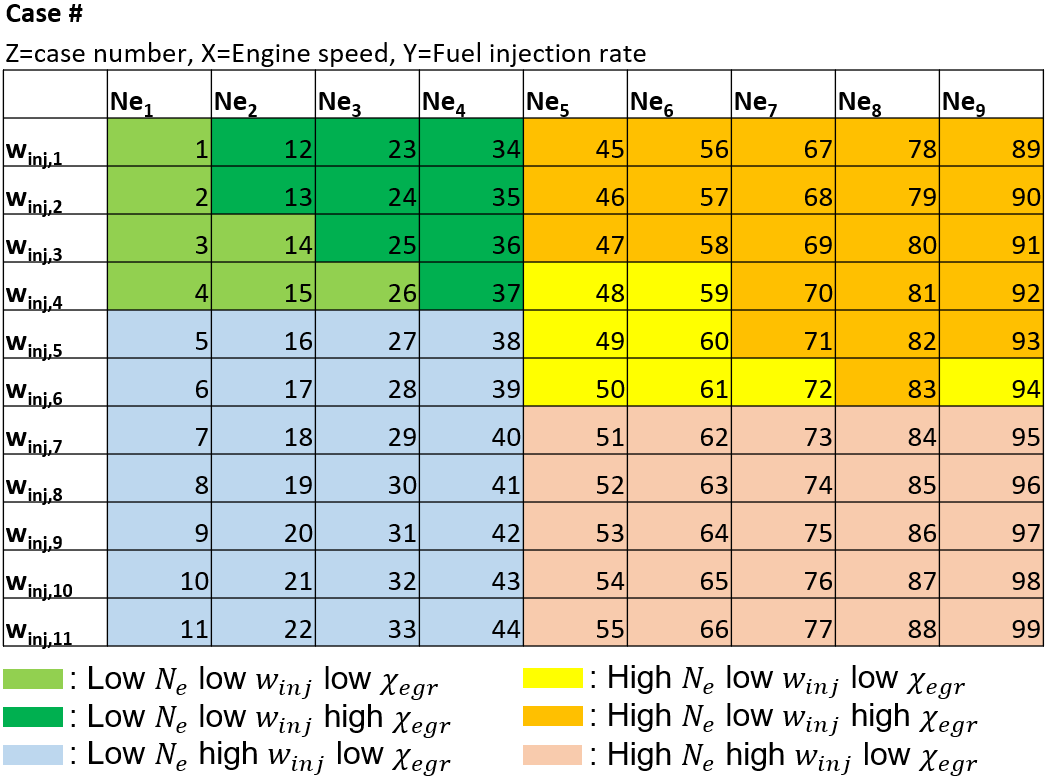}\vspace{-6pt}
\caption{Six regions defined as a function of $\chi_{egr}$, $N_e$ and $w_{inj}$ to simplify the MPC tuning process with the same $Q^{ext}$ and $R^{ext}$ pair used in each region.}\vspace{-10pt}
\label{fig:QR}
\end{figure}

\textcolor{black}{The main limitation of the tuning framework discussed above is that it needs to be repeated for each operating condition. Given there are 99 operating points, the MPC tuning process would be burdensome and time-consuming. To make the tuning process more tractable, we propose to group multiple operating points and tune the MPC for each group. The 99 engine operating points are categorized into six regions as a function of $\chi_{egr}$, $N_e$ and $w_{inj}$ as shown in Fig.~\ref{fig:QR}. In each region, the same $Q^{ext}$ and $R^{ext}$ are used for MPC implementation. The six regions are defined as follows:
\begin{itemize}
    \item low $N_e$, low $w_{inj}$, low $\chi_{egr}$,
    \item low $N_e$, low $w_{inj}$, high $\chi_{egr}$,
    \item low $N_e$, high $w_{inj}$, low $\chi_{egr}$,
    \item high $N_e$, low $w_{inj}$, low $\chi_{egr}$,
    \item high $N_e$, low $w_{inj}$, high $\chi_{egr}$,
    \item high $N_e$, high $w_{inj}$, low $\chi_{egr}$.
\end{itemize}
Note that $\chi_{egr}$, $N_e$, and $w_{inj}$ are categorized using qualitative ``high'' and ``low'' terms to protect the OEM proprietary data.}
%

\vspace{-7pt}
\section{Simulation Results and Discussion}\label{sec:results}
\textcolor{black}{Closed-loop simulations of airpath MPC \eqref{eq:modified MPC} are conducted over both step and transient drive cycle tests using the high-fidelity GT-Power model with transient thermal dynamics as the plant that provides feedback to the closed-loop MPC. A sampling and control update period of 20 $ms$ is used, the same as the prediction model step size. The prediction horizon is $N = 50$, which is chosen from the average response time of $p_{im}$ and $\chi_{egr}$ from GT-Power. The MPC is implemented using LPV \textit{Model-A}, \textit{Model-B}, and \textit{Model-C} with soft state constraints and hard control constraints. The package {\tt MPCTools} \cite{risbeck2016mpctools} is used to numerically solve the MPC optimization problem.}

\vspace{-3pt}
\subsection{Simple Reference Tracking Case Study}
\begin{figure}[h!]
    \centering
    \subfigure{\includegraphics[width=0.238\textwidth]{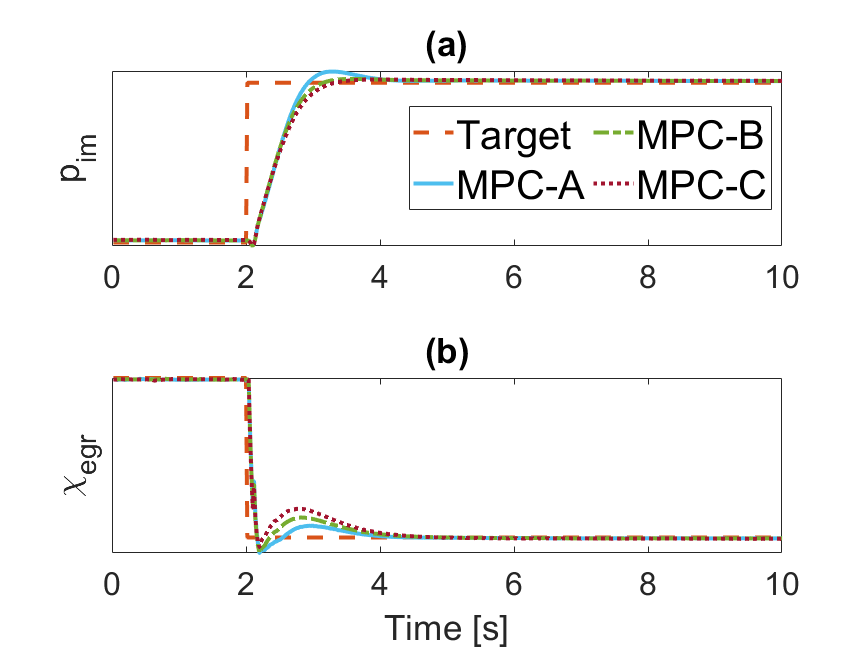}} 
    \subfigure{\includegraphics[width=0.238\textwidth]{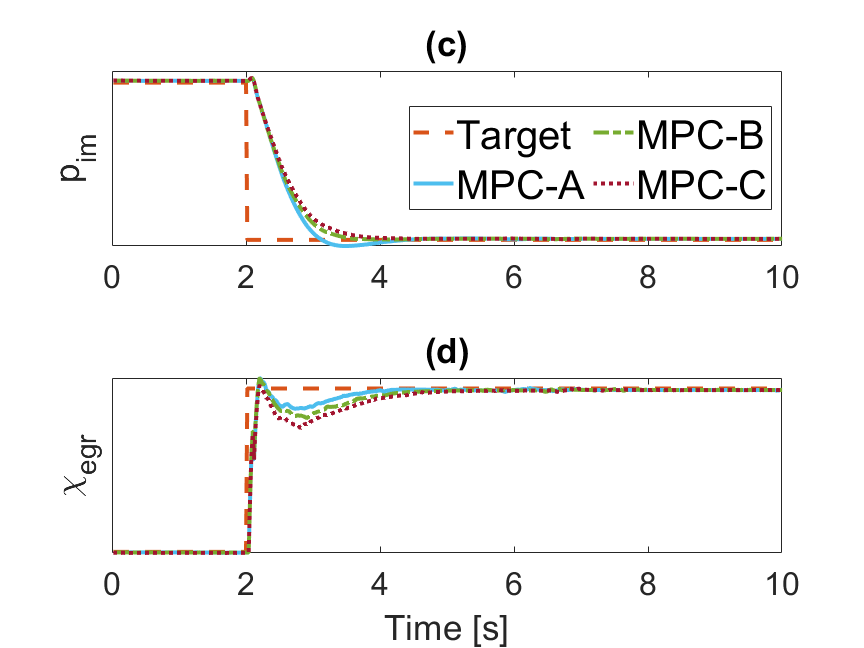}} 
    \caption{Example of (a)\&(c) $p_{im}$ and (b)\&(d) $\chi_{egr}$ trajectory tracking from \textbf{MPC-A}, \textbf{MPC-B} and \textbf{MPC-C} during a fuel tip-in (left) and tip out (right) simulation at 2000 RPM.}\vspace{-16pt}
    \label{fig:tip_in_out}
\end{figure}

\textcolor{black}{Fig.~\ref{fig:tip_in_out}} shows the comparison of $p_{im}$ and $\chi_{egr}$ response among the MPCs with LPV \emph{Model-A} (\textbf{MPC-A}), with LPV \emph{Model-B} (\textbf{MPC-B}), and with LPV \emph{Model-C} (\textbf{MPC-C}) during fuel injection rate tip-in and tip-out, respectively. All of the cases have response time less than 2 $s$ and overshoot less than 5\%. According to \textcolor{black}{Fig.~\ref{fig:tip_in_out}}, all three controllers demonstrate acceptable tracking performance. The \textbf{MPC-A} has the largest overshoot in $p_{im}$ and the smallest overshot in $\chi_{egr}$ while \emph{Model-C} has the smallest overshoot in $p_{im}$ and largest overshot in $\chi_{egr}$.~To further test the MPC performance, three controllers are tested based on transient drive cycle in the following sub-section. 
\vspace{-15pt}

\subsection{Transient Drive Cycle Implementation}
\begin{figure}[h!]
\centering
\includegraphics[width=0.95\columnwidth]{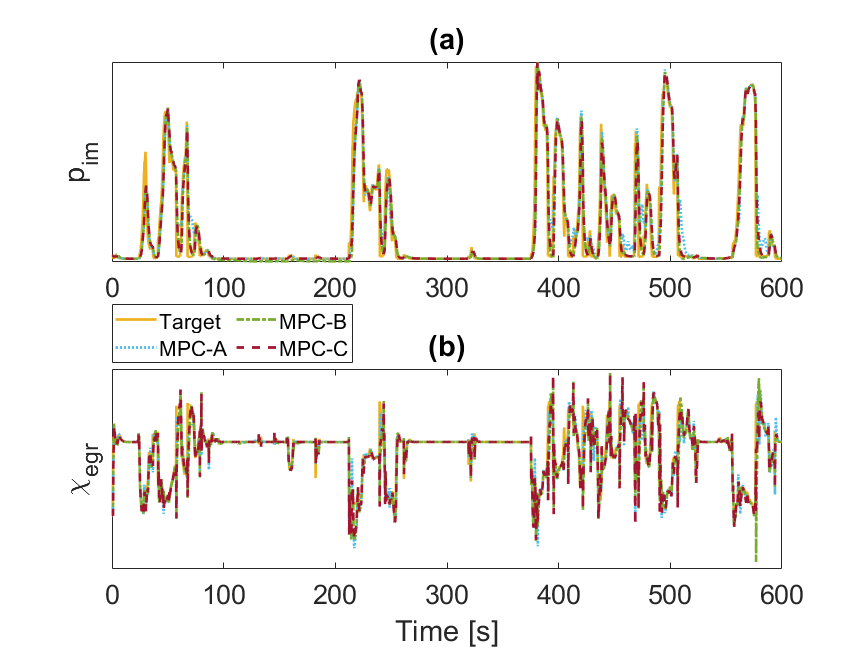}\vspace{-6pt}
\caption{(a) $p_{im}$ and (b) $\chi_{egr}$ tracking performances from \textbf{MPC-A}, \textbf{MPC-B}, and \textbf{MPC-C} over the first 600 $s$ of FTP simulation.}\vspace{-18pt}
\label{fig:FTP}
\end{figure}

\begin{figure}[h!]
\centering
\includegraphics[width=0.95\columnwidth]{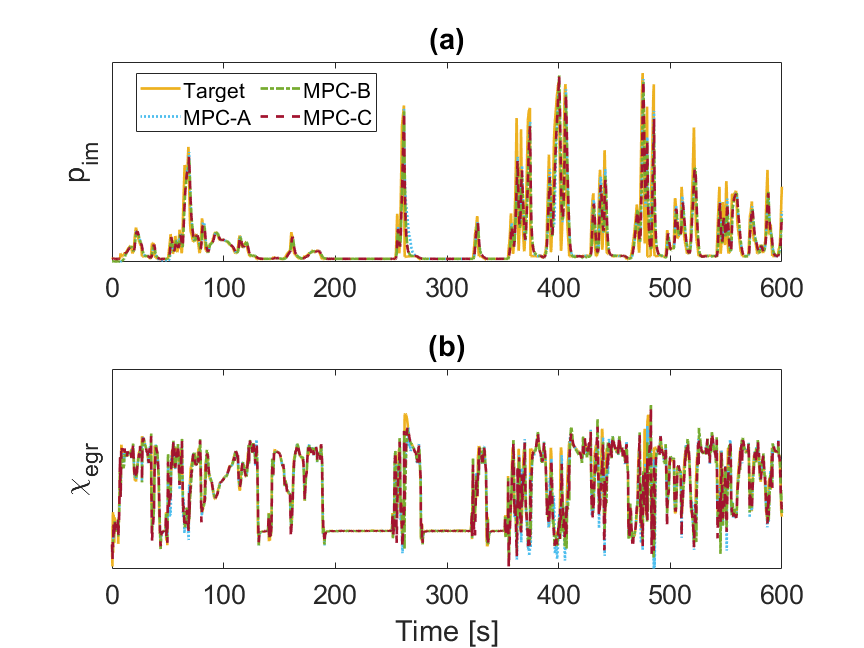}\vspace{-6pt}
\caption{(a) $p_{im}$ and (b) $\chi_{egr}$ tracking performances from \textbf{MPC-A}, \textbf{MPC-B}, and \textbf{MPC-C} over the first 600 $s$ of WHTC simulation.}\vspace{-12pt}
\label{fig:WHTC}
\end{figure}

Figs.~\ref{fig:FTP} and \ref{fig:WHTC}, and Tables~\ref{tbl:FTP} and \ref{tbl:WHTC} show the comparison of $p_{im}$ and $\chi_{egr}$ response from the \textbf{MPC-A}, \textbf{MPC-B}, and \textbf{MPC-C} over the first 600 $s$ of the FTP and World Harmonized Transient Cycle (WHTC), respectively. Compared to the baseline MPC, both \textbf{MPC-B} and \textbf{MPC-C} have better tracking performance for $p_{im}$ tracking while having a similar or slightly degraded performance for $\chi_{egr}$ tracking over FTP. Over the first 600 $s$ of WHTC, both \textbf{MPC-B} and \textbf{MPC-C} have better performance for $p_{im}$ tracking. \textbf{MPC-C} has the best performance for $\chi_{egr}$ tracking. However, the $\chi_{egr}$ tracking error with \textbf{MPC-B} is larger than that of \textbf{MPC-A}. This observation suggests that it is beneficial to consider $w_{inj}$ as an additional input to the LPV model, which also fits with the conclusion from Fig.~\ref{fig:step comparison}. Overall, all three MPCs demonstrate robustness against transient engine thermal dynamics, and are able to track the target set-points with no steady-state errors. It should be noted that all three MPCs are tuned based on the process described in \cref{sec:control implementation}. \textcolor{black}{Tables~\ref{tbl:FTP} and \ref{tbl:WHTC} also list the results over FTP and WHTC with only feed-forward controller for comparison, demonstrating the superior tracking performance of the combined feedforward and feedback controller as compared to the feed-forward case.}~
\vspace{-12pt}

\begin{table}[h!]
\caption{Comparison of the three MPCs performances over the FTP.}\vspace{-10pt}
\label{tbl:FTP}
{\scriptsize
\begin{center}
\begin{tabular}{lllll}
\toprule  
\makecell[l]{\textbf{MPC}} & \makecell[l]{\textbf{ $\overline{e}_{p_{im}}$}[bar]} & \makecell[l]{\textbf{$\overline{e}_{\chi_{egr}}$}} & \makecell[l]{\textbf{STD: $e_{p_{im}}$}[bar]} & \makecell[l]{\textbf{STD: $e_{\chi_{egr}}$}}\\
\midrule  
\textbf{MPC-A} & 0.0789 & 0.0113 & 0.1509 & 0.0240\\ 
(reference) &  &  &  &\\ \hline
\textbf{MPC-B} & 0.0705 & 0.0129 & 0.1368 & 0.0309 \\
 & ($\downarrow$ 10.6\%) & ($\uparrow$ 14.2\%) & ($\downarrow$ 9.3\%) & ($\uparrow$ 28.8\%)\\\hline
\textbf{MPC-C} & 0.0668 & 0.0113 & 0.1389 & 0.0269\\
& ($\downarrow$ 15.3\%) & ($\downarrow$ 0\%) & ($\downarrow$ 8.0\%) & ($\uparrow$ 12.1\%)\\\hline
\textcolor{black}{\textbf{FF only}} & \textcolor{black}{0.1399} & \textcolor{black}{0.0330} & \textcolor{black}{0.2743} & \textcolor{black}{0.0345}\\
\textcolor{black}{} & \textcolor{black}{($\uparrow$ 77.31\%)} & \textcolor{black}{($\uparrow$ 192.0\%)} & \textcolor{black}{($\uparrow$ 81.78\%)} & \textcolor{black}{($\uparrow$ 43.75\%)}\\
\bottomrule 
\end{tabular}\vspace{-6pt}
\end{center}}
\end{table}

\begin{table}[h!]
\caption{Comparison of the three MPCs performances over the WHTC.}\vspace{-10pt}
\label{tbl:WHTC}
{\scriptsize
\begin{center}
\begin{tabular}{lllll}
\toprule  
\makecell[l]{\textbf{MPC}} & \makecell[l]{\textbf{ $\overline{e}_{p_{im}}$}[bar]} & \makecell[l]{\textbf{$\overline{e}_{\chi_{egr}}$}} & \makecell[l]{\textbf{STD: $e_{p_{im}}$}[bar]} & \makecell[l]{\textbf{STD: $e_{\chi_{egr}}$}}\\
\midrule  
\textbf{MPC-A} & 0.0821 & 0.0127 & 0.1528 & 0.0231\\ 
(reference) &  &  &  &\\ \hline
\textbf{MPC-B} & 0.0801 & 0.0135 & 0.1502 & 0.0258\\
& ($\downarrow$ 2.4\%) & ($\uparrow$ 6.3\%) & ($\downarrow$ 1.7\%) & ($\uparrow$ 11.7\%)\\ \hline
\textbf{MPC-C} & 0.0767 & 0.0113 & 0.1453 & 0.0215\\
& ($\downarrow$ 6.6\%) & ($\downarrow$ 11\%) & ($\downarrow$ 4.9\%) & ($\downarrow$ 6.9\%)\\\hline
\textcolor{black}{\textbf{FF only}} & \textcolor{black}{0.1020} & \textcolor{black}{0.0558} & \textcolor{black}{0.2121} & \textcolor{black}{0.0434}\\
\textcolor{black}{} & \textcolor{black}{($\uparrow$ 24.24\%)} & \textcolor{black}{($\uparrow$ 339.4\%)} & \textcolor{black}{($\uparrow$ 38.81\%)} & \textcolor{black}{($\uparrow$ 87.88\%)}\\
\bottomrule 
\end{tabular}\vspace{-20pt}
\end{center}}
\end{table}

\section{Summary and Conclusions}\label{sec:conclusion}
This paper described the development of a model predictive controller (MPC) for diesel engine airpath coordinated control by EGR valve and VGT actuators. A high-fidelity model of the engine in GT-Power was used as an engine surrogate for controller development and verification. By identifying the LPV prediction model from GT-Power with transient thermal solver and adding fuel injection rate to be the third input of the model, it was shown that the tracking performance of the MPC could be improved compared to when the model is identified from steady-state thermal data and has two inputs.

\vspace{-3pt}
\section*{Acknowledgment}
Dominic Liao-McPherson from ETH Zurich is gratefully acknowledged for the technical comments provided during the course of this study.

\bibliographystyle{IEEEtran}

\bibliography{IEEEabrv,ACC2022_MPCDiesel}

\begin{thebibliography}{10}
\providecommand{\url}[1]{#1}
\csname url@samestyle\endcsname
\providecommand{\newblock}{\relax}
\providecommand{\bibinfo}[2]{#2}
\providecommand{\BIBentrySTDinterwordspacing}{\spaceskip=0pt\relax}
\providecommand{\BIBentryALTinterwordstretchfactor}{4}
\providecommand{\BIBentryALTinterwordspacing}{\spaceskip=\fontdimen2\font plus
\BIBentryALTinterwordstretchfactor\fontdimen3\font minus
  \fontdimen4\font\relax}
\providecommand{\BIBforeignlanguage}[2]{{%
\expandafter\ifx\csname l@#1\endcsname\relax
\typeout{** WARNING: IEEEtran.bst: No hyphenation pattern has been}%
\typeout{** loaded for the language `#1'. Using the pattern for}%
\typeout{** the default language instead.}%
\else
\language=\csname l@#1\endcsname
\fi
#2}}
\providecommand{\BIBdecl}{\relax}
\BIBdecl

\bibitem{del2010automotive}
L.~Del~Re, F.~Allg{\"o}wer, L.~Glielmo, C.~Guardiola, and I.~Kolmanovsky,
  \emph{Automotive model predictive control: models, methods and
  applications}.\hskip 1em plus 0.5em minus 0.4em\relax Springer, 2010, vol.
  402.

\bibitem{ortner2007predictive}
P.~Ortner and L.~Del~Re, ``Predictive control of a diesel engine air path,''
  \emph{IEEE Transactions on Control Systems Technology}, vol.~15, no.~3, pp.
  449--456, 2007.

\bibitem{stewart2008model}
G.~Stewart and F.~Borrelli, ``A model predictive control framework for
  industrial turbodiesel engine control,'' in \emph{47th IEEE Conference on
  Decision and Control (CDC)}, 2008, pp. 5704--5711.

\bibitem{huang2013towards}
M.~Huang, H.~Nakada, S.~Polavarapu, R.~Choroszucha, K.~Butts, and
  I.~Kolmanovsky, ``Towards combining nonlinear and predictive control of
  diesel engines,'' in \emph{2013 American Control Conference (ACC)}, 2013, pp.
  2846--2853.

\bibitem{moriyasu2019diesel}
R.~Moriyasu, S.~Nojiri, A.~Matsunaga, T.~Nakamura, and T.~Jimbo, ``Diesel
  engine air path control based on neural approximation of nonlinear {MPC},''
  \emph{Control Engineering Practice}, vol.~91, p. 104114, 2019.

\bibitem{huang2015nonlinear}
M.~Huang, H.~Nakada, K.~Butts, and I.~Kolmanovsky, ``Nonlinear model predictive
  control of a diesel engine air path: A comparison of constraint handling and
  computational strategies,'' \emph{5th IFAC Conference on Nonlinear Model
  Predictive Control (NMPC)}, vol.~48, no.~23, pp. 372--379, 2015.

\bibitem{liao2020model}
D.~Liao-McPherson, M.~Huang, S.~Kim, M.~Shimada, K.~Butts, and I.~Kolmanovsky,
  ``Model predictive emissions control of a diesel engine airpath: Design and
  experimental evaluation,'' \emph{International Journal of Robust and
  Nonlinear Control}, vol.~30, no.~17, pp. 7446--7477, 2020.

\bibitem{GTmanual}
\text{Gamma} Technologies, \emph{GT-SUITE Help}, 601 Oakmont Ln, Westmont, IL,
  2020.

\bibitem{huang2018toward}
M.~Huang, D.~Liao-McPherson, S.~Kim, K.~Butts, and I.~Kolmanovsky, ``Toward
  real-time automotive model predictive control: A perspective from a diesel
  air path control development,'' in \emph{2018 Annual American Control
  Conference (ACC)}, 2018, pp. 2425--2430.

\bibitem{wang2004tutorial}
L.~Wang, ``A tutorial on model predictive control: Using a linear velocity-form
  model,'' \emph{Developments in Chemical Engineering and Mineral Processing},
  vol.~12, no. 5-6, pp. 573--614, 2004.

\bibitem{pannocchia2015offset}
G.~Pannocchia, M.~Gabiccini, and A.~Artoni, ``Offset-free {MPC} explained:
  Novelties, subtleties, and applications,'' \emph{5th IFAC Conference on
  Nonlinear Model Predictive Control (NMPC)}, vol.~48, no.~23, pp. 342--351,
  2015.

\bibitem{risbeck2016mpctools}
M.~Risbeck and J.~Rawlings, ``{MPCTools: Nonlinear Model Predictive Control
  Tools for CasADi},'' 2016, {[online] Available:
  \tt\url{https://bitbucket.org/rawlings-group/octave-mpctools}}.

\end{thebibliography}

\end{document}